\documentstyle[12pt,epsf]{article}

\def\Ms{M_{\odot}}

\def\Mpc{{\rm Mpc}}
\def\kpc{{\rm kpc}}
\def\pc{{\rm pc}}
 
\def\etal{{\frenchspacing\it et al.}}
\def\ie{{\frenchspacing\it i.e.}}
\def\eg{{\frenchspacing\it e.g.}}

\def\crr{\cr\noalign{\vskip 4pt}}
\def\pp{\noindent\parshape 2 0truecm 15.6truecm 2truecm 13.6truecm}
\def\rf#1;#2;#3;#4 {\par\pp#1, {\it #2}, {\bf #3}, #4. \par}
\def\rn{\pp}
 
\def\erfc{{\rm erfc}}
\def\st{\sigma_t}
\def\Ob{\Omega_b}
\def\Oz{\Omega_0}
\def\Lz{\lambda_0}

\def\taut{\tau}
\def\izi{\int_0^{\infty}}

\def\izz{\int_0^z}

\def\fupp{f_{uvpp}}

\def\fs{f_s}

\def\fion{f_{ion}}

\def\zvir{z_{vir}}
\def\zion{z_{ion}}
\def\boost{B}
\def\v{\chi}     

\def\beq#1{\begin{equation}\label{#1}}
\def\eeq{\end{equation}}
\def\beqa#1{\begin{eqnarray}\label{#1}}
\def\eeqa{\end{eqnarray}}
\def\eq#1{equation~(\ref{#1})}
\def\eqnum#1{~(\ref{#1})}

\def\spose#1{\hbox to 0pt{#1\hss}}
\def\simlt{\mathrel{\spose{\lower 3pt\hbox{$\mathchar"218$}}
     \raise 2.0pt\hbox{$\mathchar"13C$}}}
\def\simgt{\mathrel{\spose{\lower 3pt\hbox{$\mathchar"218$}}
     \raise 2.0pt\hbox{$\mathchar"13E$}}}
\def\simpropto{\mathrel{\spose{\lower 3pt\hbox{$\mathchar"218$}}
     \raise 2.0pt\hbox{$\propto$}}}

\def\addr#1{{\small\it #1}}
\def\auth#1{{#1}}
 
\begin{document}


\begin{titlepage}   

\begin{center}

\vskip0.9truecm
{\bf

REIONIZATION IN AN OPEN CDM UNIVERSE:

IMPLICATIONS FOR

COSMIC MICROWAVE BACKGROUND FLUCTUATIONS\footnote{
Published in {\it ApJ}, {\bf 441}, 458, March 10, 1995.\\
Submitted May 3 1994, accepted September 13.
Available from\\
{\it h t t p://www.sns.ias.edu/$\tilde{~}$max/openreion.html} 
(faster from the US) and from\\
{\it h t t p://www.mpa-garching.mpg.de/$\tilde{~}$max/openreion.html} 
(faster from Europe).\\
}
}

\vskip 0.5truecm
  \auth{Max Tegmark}

  \smallskip
  \addr{Department of Physics, University of California,
Berkeley, California  94720}
  \smallskip
  \vskip 0.2truecm

  \auth{Joseph Silk}

  \smallskip
  \addr{Departments of Astronomy and Physics, and
Center for Particle Astrophysics,}
 
  \addr{University of California, Berkeley, California 94720}
  \smallskip

\end{center}
 
 
\def\taut{\tau}

\abstract{
We generalize previous work on early photoionization 
to CDM models with $\Omega<1$.
Such models have received recent interest because  
the excess power in the large--scale galaxy distribution 
is phenomenologically fit if the ``shape parameter"
$\Gamma \equiv h\Omega_0\approx 0.25.$
It has been argued that such models may require early
reionization to suppress small-scale anisotropies
in order to be consistent with experimental data.
We find
that if
the cosmological constant  $\lambda = 0$, the extent of this suppression is
quite insensitive to $\Omega_0$.
Given a $\sigma_8$-normalization today, 
the loss of small--scale power associated
with a lower $\Omega_0$ is partially canceled by higher optical depth from
longer lookback times and by structures forming at higher
redshifts before
the universe becomes curvature--dominated.
The maximum angular scale on which fluctuations are
suppressed decreases when $\Omega_0$ is lowered,
but this effect is also rather weak and unlikely to be
measurable in the near future.
For flat models, on the other hand, where
$\lambda_0 = 1 - \Omega_0$,  the negative effects of lowering $\Omega_0$
dominate,  and early reionization is not likely to play a
significant role if $\Omega_0\ll 1$. The same applies to 
CDM models 
where the effective $\Gamma$ is lowered by increasing the number of
relativistic particle species.
}

\end{titlepage}


\section{Introduction}

Early reionization of the universe would affect fluctuations in the cosmic microwave background radiation (CBR) on degree-scales and sub--degree scales. Considerable effort is now being devoted to measuring such fluctuations, so a detailed
knowledge of the ionization history is desirable in order to give the experimental results their proper cosmological interpretation.

It is generally believed that the intergalactic medium (IGM) was reionized at some time in the past. The main reason for this is the absence of a Gunn-Peterson trough in the spectra of high redshift quasars (Gunn \& Peterson 1965; Steidel \& Sargent 1987; Webb {\etal} 1992), indicating that reionization occurred at least as early as $z=4$. For CBR applications, the relevant question is not whether reionization occurred, but when it occurred. 
For instance, to 
provide an optical depth to scattering exceeding $20\%$,
which suffices to reconcile CDM with all observational CBR limits (Sugiyama, Silk \& Vittorio 1993),
reionization must have occurred by redshift $z=30$, adopting 
the nucleosynthesis value of the baryon density. 

In Tegmark, Silk \& Blanchard 1994 (``Paper 1"),  
early
photoionization of the intergalactic medium was discussed in a fairly
model--independent way, in order to  investigate whether early structures
corresponding to rare Gaussian peaks in 
a CDM model could photoionize the
intergalactic 
medium sufficiently early to appreciably smooth out the microwave
background fluctuations. In this paper, the results of 
Paper 1
will be
generalized to $\Omega<1$ models with non--zero cosmological constant
$\Lambda$.  Essentially all the notation used in this paper was defined in
Paper 1 and,
in the interest of brevity, some of the definitions will not be
repeated here.

Just as in Paper 1,
our basic picture is the following: 
An ever larger fraction $\fs$ of the baryons in the universe 
falls into nonlinear
structures and forms galaxies. A certain fraction of these baryons form
stars or quasars which emit ultraviolet radiation. 
Some of this radiation escapes into the ambient 
intergalactic medium (IGM), which is consequently  
photoionized and heated. Due to cooling losses and
recombinations, the net number of ionizations per UV
photon, $\fion$, is generally less than unity. 

The results that we present here generalize the previous 
work to the case of an
open universe.
Lowering $\Oz$ has four distinct effects: 

\begin{enumerate}

\item
Density fluctuations gradually stop growing once 
$z\simlt\Oz^{-1}$. 
Thus given the observed power spectrum today, a lower $\Oz$ implies that 
the first structures formed earlier.

\item
Matter-radiation equality occurs later, which shifts the
turning-point of the CDM power spectrum toward larger scales.
This means less power on very small scales (such as $\sim 10^6\Ms$) relative to the
scales at which we normalize the power spectrum (namely galaxy cluster scales, $\sim 8h^{-1}\Mpc$ or even  
the much larger COBE scale). One consequence is that the first structures form later.

\item
The lookback time to a given ionization redshift becomes larger,
resulting in a higher optical depth.

\item
The horizon at a given ionization redshift subtends a
smaller angle on the sky, thus lowering the angular scale 
below which CBR fluctuations are suppressed.

\end{enumerate}

\noindent
Thus in terms of the virialization redshift $\zvir$ defined in 
Paper 1,
the
redshift at which typical structures to go nonlinear, 
effect~1 increases $\zvir$ whereas effect~2 decreases $\zvir$.
These two effects influence $\fs$, the fraction of baryons in nonlinear
structures, in opposite directions.
As to effect~2, it should be noted that this applies not only to CDM, but to
any spectrum that ``turns over" somewhere between the very smallest nonlinear
scales  ($\sim 10^6\Ms$) and the very largest  
($\sim 10^{21}\Ms$)
scales at which we COBE-normalize. 
Yet another effect of lowering $\Oz$ is that 
the ionization efficiency $\fion$ drops slightly, at most by
a factor $\Oz^{1/2}$. This is completely negligible compared to the
above-mentioned effects, as $\zion$ depends only logarithmically on the
efficiency. 

In the following sections, we will 
discuss each of these four effects in greater detail,  
and then compute their
combined modification of the ionization history in a few scenarios.

\section{The Boost Factor}

\begin{figure}[phbt]
\centerline{\epsfxsize=17cm\epsfbox{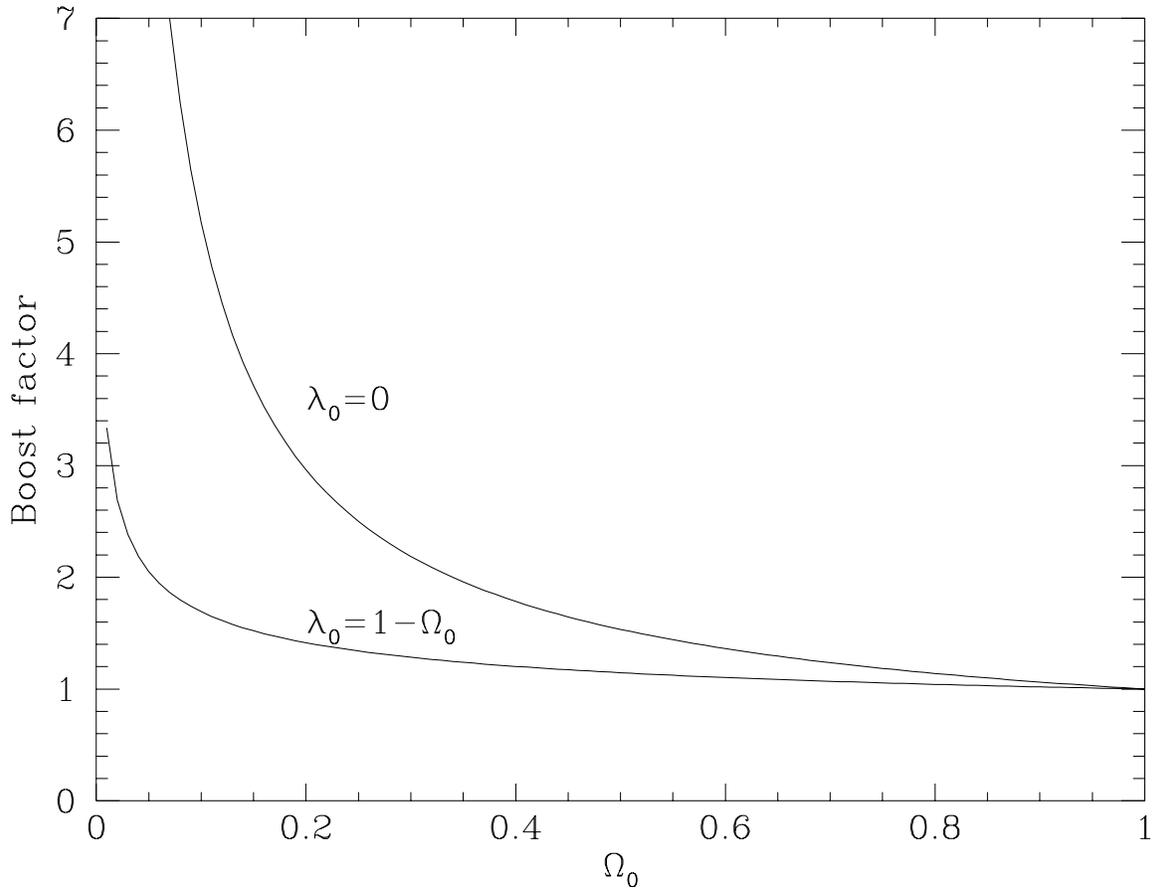}}
\caption{The boost factor.}
\label{openreionfig1}
The boost factor $B(\Omega_0,\lambda_0)$ is plotted as a function 
of $\Omega_0$ for two classes of cosmologies.
The upper curve corresponds to a standard open universe, 
{\it i.e.} $\lambda_0=0$, whereas the lower curve corresponds to flat
universes with 
$\lambda_0=1-\Omega_0$.
\end{figure}

When curvature and vacuum density are negligible, 
sub-horizon-sized density fluctuations simply grow as the
scale factor $a\propto (1+z)^{-1}$. 
Thus at early times $z\gg\Oz^{-1}$, we can write 
$$\delta =  {\boost(\Oz,\Lz)\over 1+z}\delta_0$$
for some function $\boost$ independent of $z$ that we will refer to as the
{\it boost factor}. Clearly $\boost(1,0) = 1$.
Thus if certain structures are assumed to form when $\delta$ equals some fixed
value, then given the observed power spectrum today, the boost factor tells us
how much earlier these structures would form than they would in a standard flat
universe. 
The boost factor is simply the inverse of the so called
growth factor, and can be computed analytically for 
a number of special cases (see {\eg} Peebles 1980).
For the most general case, the fit 
$$
B(\Oz,\Lz) \approx
{2\over 5\Oz}\left[\Oz^{4/7} - \Lz + 
\left(1+{\Oz\over 2}\right)\left(1+{\Lz\over 70}\right)\right]
$$ 
is accurate to within a few percent for all parameter values of cosmological
interest
(Carroll {\etal} 1992).
The exact results are plotted in 
Figure~\ref{openreionfig1} 
for the case $\Lz=0$
and the flat case $\Lz = 1-\Omega_0$.
Since we will limit ourselves to these two cases,
the simple power-law fits
$$\cases{
\boost(\Oz,0) & $\approx\Oz^{-0.63},$\crr
\boost(\Oz,1-\Oz) & $\approx\Oz^{-0.21},$
}$$
which are accurate to within $1\%$ for $0.2\leq\Oz\leq 1$,
will suffice for our purposes. 
Note that the standard rule of thumb that
perturbations stop growing at $1+z\approx\Oz^{-1}$, indicating 
$B\propto\Oz^{-1}$, is not particularly accurate in this context.

\section{The Power Spectrum Shift}

As mentioned above, lowering $h\Oz$ causes the first structures go 
nonlinear at a later redshift. This is quantified in the present section.

The standard 
CDM model with power-law initial fluctuations 
proportional to
$k^n$ predicts a power
spectrum that is well fitted by (Bond \& Efstathiou 1984;
Efstathiou {\etal} 1992)
$$P(k) \propto 
{q^n\over
\left(1+\left[aq+(bq)^{1.5} +
(cq)^2\right]^{1.13}\right)^{2/1.13}},$$
where $a\equiv 6.4$, $b\equiv 3.0$, $c\equiv 1.7$, 
$q\equiv (1h^{-1}\Mpc) k/\Gamma$ and the ``shape parameter" 
$\Gamma$
will be discussed below.
Although this fit breaks down for scales comparable to the 
curvature scale  $r_{curv} = H_0^{-1}|1-\Oz|^{-1/2}$,
it is quite accurate for the much smaller scales that will be considered in
the present paper. Rather, its main limitation is that it breaks down if 
$\Oz$ is so low that the baryon density becomes comparable to the 
density of cold dark matter. Thus for $\Ob\approx 0.05$,
the results cannot be taken too seriously for $\Oz<0.2$.
We will limit ourselves to the standard $n=1$ model here, 
as the tilted ($n<1$) case was treated in 
Paper 1
and was seen
to be be essentially unable 
to reionize the universe early enough 
to be relevant to CBR anisotropies.  
The same applies to models with mixed hot and cold dark matter.

\begin{figure}[phbt]
\centerline{\epsfxsize=17cm\epsfbox{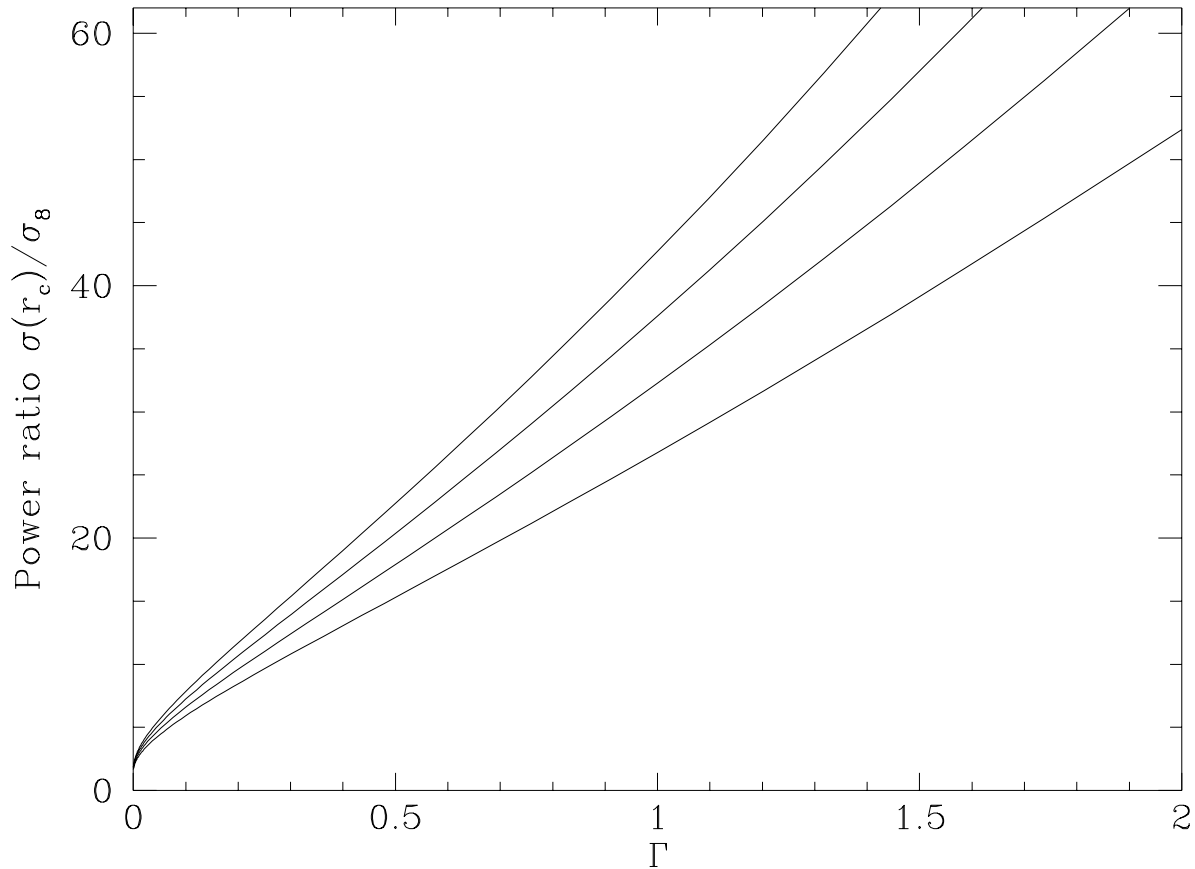}}
\caption{The amplitude ratio.}
\label{openreionfig2}
The ratio of the fluctuation 
amplitude on the small scale $r_c$ to that at 
$8h^{-1}$Mpc is plotted as a function of the shape parameter $\Gamma$.
From top to bottom, the four curves correspond to scales of 
$3.5 h^{-1}$kpc,  
$7.5 h^{-1}$kpc,  
$16 h^{-1}$kpc and 
$35 h^{-1}$kpc, respectively.
For $h=0.5$ and $\Omega_0=1$, these four length scales correspond to
the masses 
$10^5M_{\odot}$,
$10^6M_{\odot}$,
$10^7M_{\odot}$ and
$10^8M_{\odot}$.
The weak additional dependence on $\Omega_0/h$ that would result from holding 
$M_c$ rather than $r_c$ fixed is clearly negligible, since as can be seen,
$M_c$ must vary by an entire order of magnitude to offset a mere
$20\%$ change in $\Gamma$. 
\end{figure}

Let us define the {\it amplitude ratio} 
$$R(\Gamma,r_1,r_2) \equiv {\sigma(r_1)\over\sigma(r_2)},$$
where $\sigma(r_1)$ and $\sigma(r_2)$ are the {r.m.s.} 
mass fluctuation amplitudes in
spheres of radii $r_1$ and $r_2$, {\ie}
$$\sigma(r)^2 \propto
\izi P(k) 
\left[{\sin kr\over(kr)^3} - {\cos
kr\over(kr)^2}\right]^2 dk.$$ 
As in Paper 1,
we normalize the power spectrum so that
$\sigma(8 h^{-1}\Mpc)$ equals some constant denoted by $\sigma_8$,
and 
$M_c$ will denote the characteristic 
mass of the first galaxies
to form. The corresponding comoving length scale $r_c$ is 
given by $M_c = {4\over 3}\pi r_c^3\rho$, where $\rho$ is the mean density of
the universe. 
Thus given $\sigma_8$, 
what is relevant for
determining when the first galaxies form is the amplitude ratio
$$R(\Gamma, r_c, 8 h^{-1}\Mpc).$$
This ratio is computed numerically, and the results are 
plotted as a function of $\Gamma$ in 
Figure~\ref{openreionfig2} 
for a few different 
values of the cutoff mass $M_c$. 
It is easy to see why the amplitude ratio increases with $\Gamma$, since
on a logarithmic scale, a decrease in $\Gamma$ simply shifts the entire 
power spectrum towards lower $k$, thus decreasing the amount of power on very
small scales relative to that on large scales. The fit 
$$R(\Gamma, r_c, 8 h^{-1}\Mpc)
\approx 3 + 7.1\ln(1 h^{-1}\Mpc/r_c)\Gamma$$
is accurate to within $10\%$ for $0.05<\Gamma<2$ and 
$100\pc<r_c<100\kpc$.

\section{The Optical Depth}

Since a lower $\Oz$ implies a larger $|dt/dz|$ and an older universe, 
the optical depth out to a given ionization redshift $\zion$
is greater for small $\Oz$.
For a given ionization history $\v(z)$, the 
optical depth for Thomson scattering is given by
$$\cases{
\taut(z)& = $\taut^*\izz{(1+z')^2\over
\sqrt{\Lz + (1+z')^2(1-\Lz+\Oz z')}}\v(z')dz',$\crr
\taut^*& = ${3\Ob\over 8\pi}
\left[1 - \left(1-{\v_{He}\over 4\v}\right)f_{He}\right] 
{H_0c\st\over m_p G} \approx 0.057 h\Ob,$
}$$
where we have taken the mass fraction of helium to be 
$f_{He}\approx 24\%$ and assumed $\v_{He} \approx\v$,
{\ie} that helium never becomes doubly ionized and that the fraction
that is singly ionized equals the fraction of hydrogen that is
ionized. The latter is a very crude approximation, but has the advantage that
the error can never exceed $6\%$. 
If the universe is fully ionized for all redshifts below $z$, the 
integral can be done analytically for $\Lz=0$:
\beq{tauEq}
\taut(z) = 
{2\taut^*\over 3\Oz^2}
\left[2-3\Oz + (\Oz z+3 \Oz-2)\sqrt{1+\Oz z}\,\right]
\approx 0.038{h\Ob z^{3/2}\over\Oz^{1/2}}
\eeq
for $z\gg\Oz^{-1}$. 
As is evident from the asymptotic behavior of the integrand, 
$\taut$ is independent of $\Lz$ in the high redshift limit. 
Thus optical depth of unity is attained if reionization occurs at 
$$z\approx 92 \left(h\Ob\over 0.03\right)^{-2/3}\Oz^{1/3}.$$

Besides fluctuation suppression, polarization of the CBR is another interesting probe of reionization (Crittenden, Davis \& Steinhardt 1994). Also for this application, the optical depth is a key parameter.

\section{The Angular Scale}

\begin{figure}[phbt]
\centerline{\epsfxsize=17cm\epsfbox{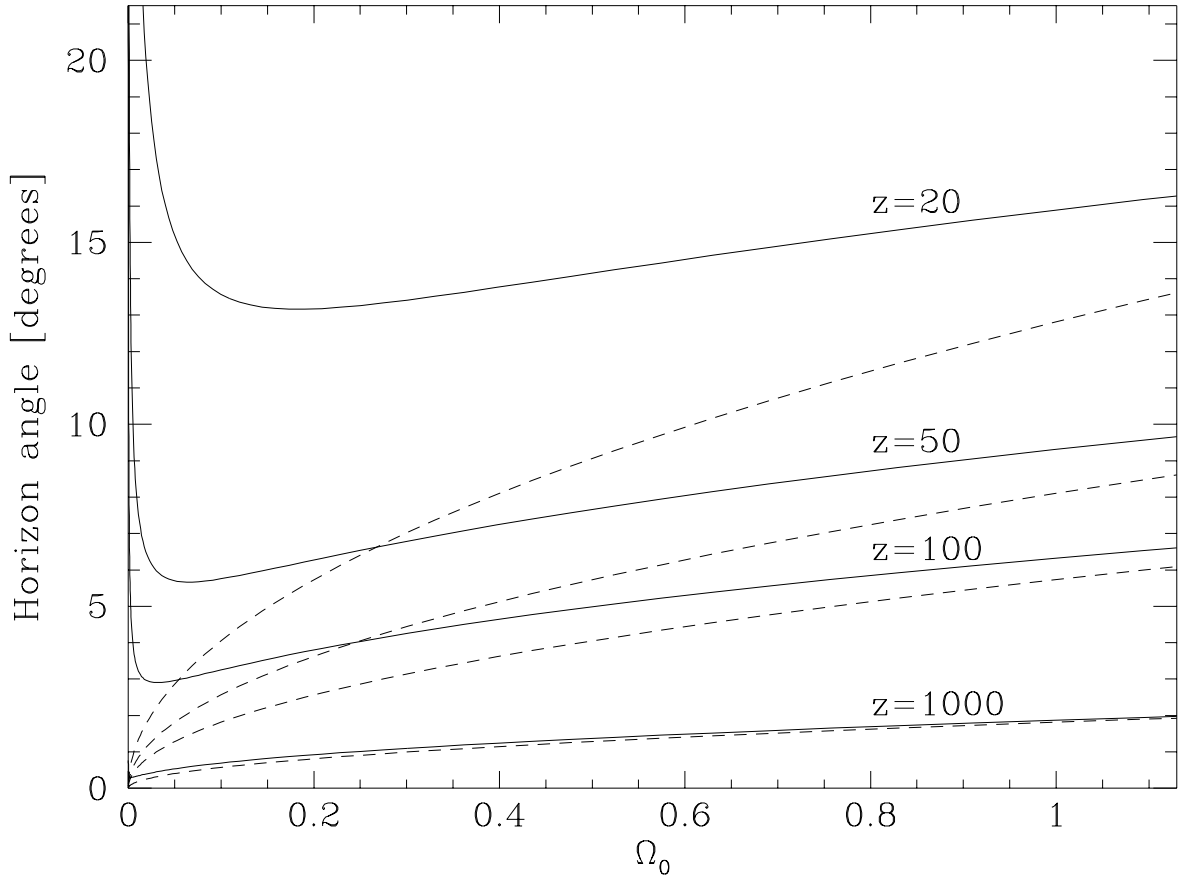}}
\caption{The horizon angle.}
\label{openreionfig3}
The angle in the sky subtended 
by a horizon volume at redshift $z$ is 
plotted as a function of $\Omega_0$ for the case with no cosmological 
constant.
The solid lines are the exact results for the four redshifts indicated,
and the dashed lines are the corresponding fits using the 
simplistic approximation $\theta\approx(\Omega_0/z)^{1/2}$.
\end{figure}

\begin{figure}[phbt]
\centerline{\epsfxsize=17cm\epsfbox{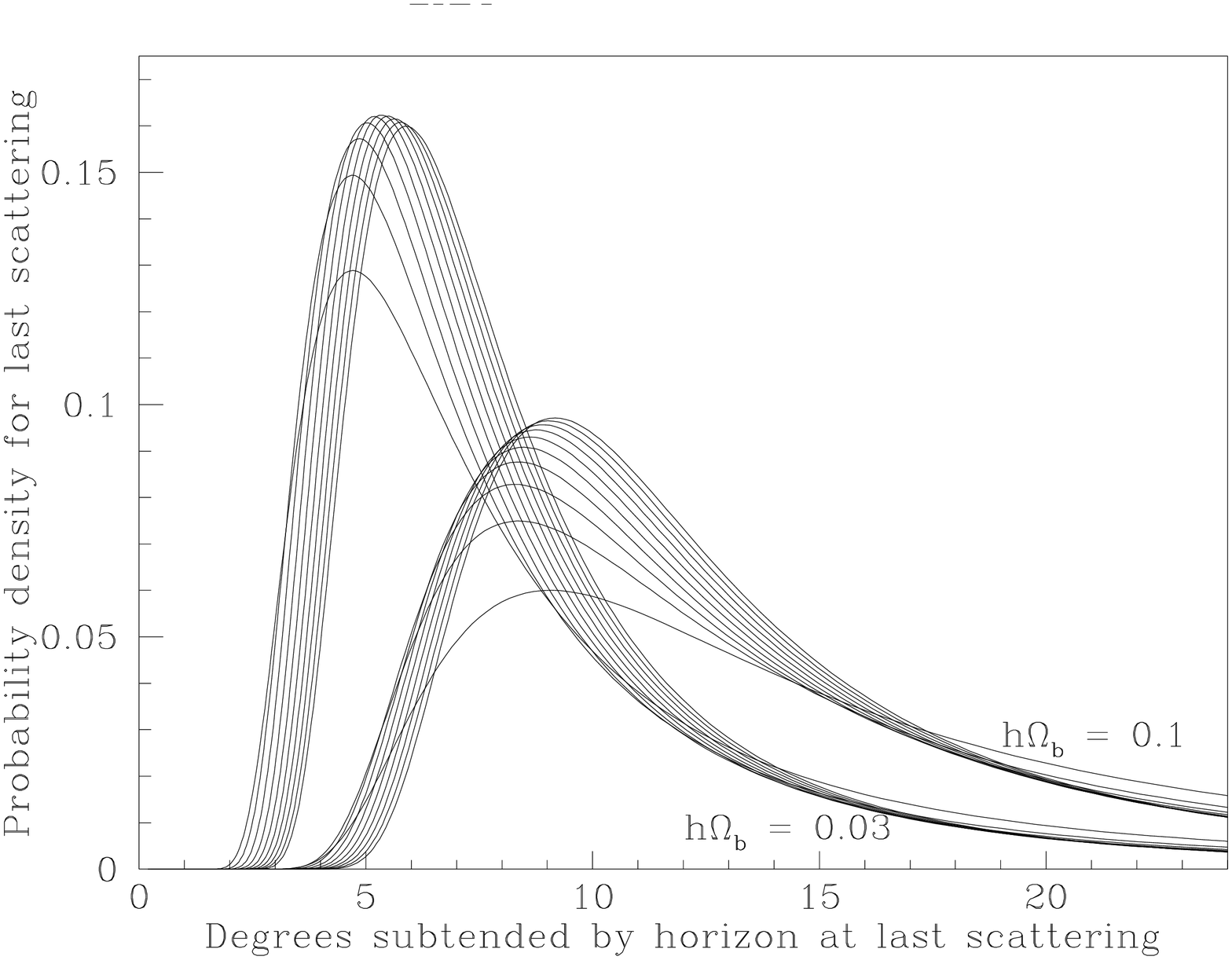}}
\caption{Visibility functions.}
\label{openreionfig4}
The angular visibility function for 
a fully ionized universe is 
plotted for different values of $\Omega_0$ and
diffuse baryon content $h\Ob$.
The left group of curves corresponds to $h\Ob = 0.03$ and
the right to $h\Ob = 0.1$.
Within each group, from left to right starting at the lowest
peak,
$\Omega_0 = $0.1, 0.2, 0.3, 0.4, 0.5, 0.6, 0.7, 0.8, 0.9 and 1.0.
\end{figure}

It is well known that reionization suppresses CBR fluctuations only on angular
scales below the horizon scale at last scattering.
Combining the standard expressions for
horizon radius ({\eg} Kolb \& Turner 1990) and angular size,
this angle is given by
\beq{thetaEq}
\theta = 2\tan^{-1}\left[
{sqrt{1+z}\Oz^{3/2}/2\over
\Oz z-(2-\Oz)(\sqrt{1+\Oz z}-1)}
\right].
\eeq
For $z\gg \Oz^{-1}$, this reduces to
\beq{thetaApproxEq}
\theta\approx{\sqrt{\Oz\over z}},
\eeq
but as is evident from
Figure~\ref{openreionfig3}, 
this is quite a bad
approximation
except for $z\gg 100$. If we substitute it into  
\eq{tauEq} nonetheless, to get a rough estimate, we
conclude that optical depth unity is obtained at an epoch 
whose horizon scale subtends the angle 
\beq{thetaEqThree}
\theta \approx 12^{\circ} \left({h\Ob\Oz\over 0.03}\right)^{1/3},
\eeq
{\ie}, the dependence on all three of these cosmological parameters is
relatively weak.
As discussed in Paper 1,
fluctuations on angular scales much smaller than this
are suppressed by a factor 
$$P(z)\equiv 1-e^{-\tau(z)},$$
the {\it opacity}, which is
the probability that a photon was Thompson scattered after redshift $z$.
Its derivative, the {\it visibility function}
$f_z = dP/dz$, is the probability distribution for the redshift at which last
scattering occurred, the profile of the last scattering surface.  
The {\it angular visibility function}
$$f_{\theta}(\theta) = \left|{dP_s\over d\theta}\right| = 
\left|{d\theta\over dz}\right|^{-1}{dP_s\over dz}$$ 
is plotted in 
Figure~\ref{openreionfig4} 
for the case where the universe never recombines 
(the curves for the more general case with reionization at
some redshift $\zion$ can be read off from 
Figure~\ref{openreionfig4} 
as described in Paper 1).
These functions give a good idea of the
range of angular scales on which suppression starts to become important.
In plotting these curves, the exact expression\eqnum{thetaEq}
has been used, rather
than the approximation\eqnum{thetaApproxEq}.
It is seen that the qualitative behavior indicated by 
\eq{thetaEqThree} is
correct: as $\Oz$ is lowered, the peak shifts down toward smaller 
angular scales,
but the $\Oz$-dependence is quite weak.

\section{Cosmological Consequences}

We will now compare the effect of lowering $\Gamma$ in three 
cosmological models. The first model, which will be referred to as 
``open CDM" for short, has $\lambda_0 = 0$.
The second model, referred to as ``$\Lambda$CDM", has 
$\lambda_0 = 1-\Oz$. 
The {\it shape parameter} essentially tells us how early the
epoch of matter-radiation equality occurred, and 
is given by
$$\Gamma = h\Oz\left({g_*\over 3.36}\right)^{-1/2},$$
where $g_*	= 3.36$ corresponds to the standard model with
no other relativistic degrees of freedom than photons and 
three massless neutrino species. 
In open CDM and  $\Lambda$CDM, we have 
$g_* = 3.36$, so that $\Omega_0 = \Gamma/h$. 
The third model, referred to as $\tau$CDM
(Dodelson, Gyuk \& Turner 1994), has 
$\lambda_0=0$ and $\Omega = 1$, and achieves a lower value of $\Gamma$ 
by increasing $g_*$ instead. 

\begin{figure}[phbt]
\centerline{\epsfxsize=17cm\epsfbox{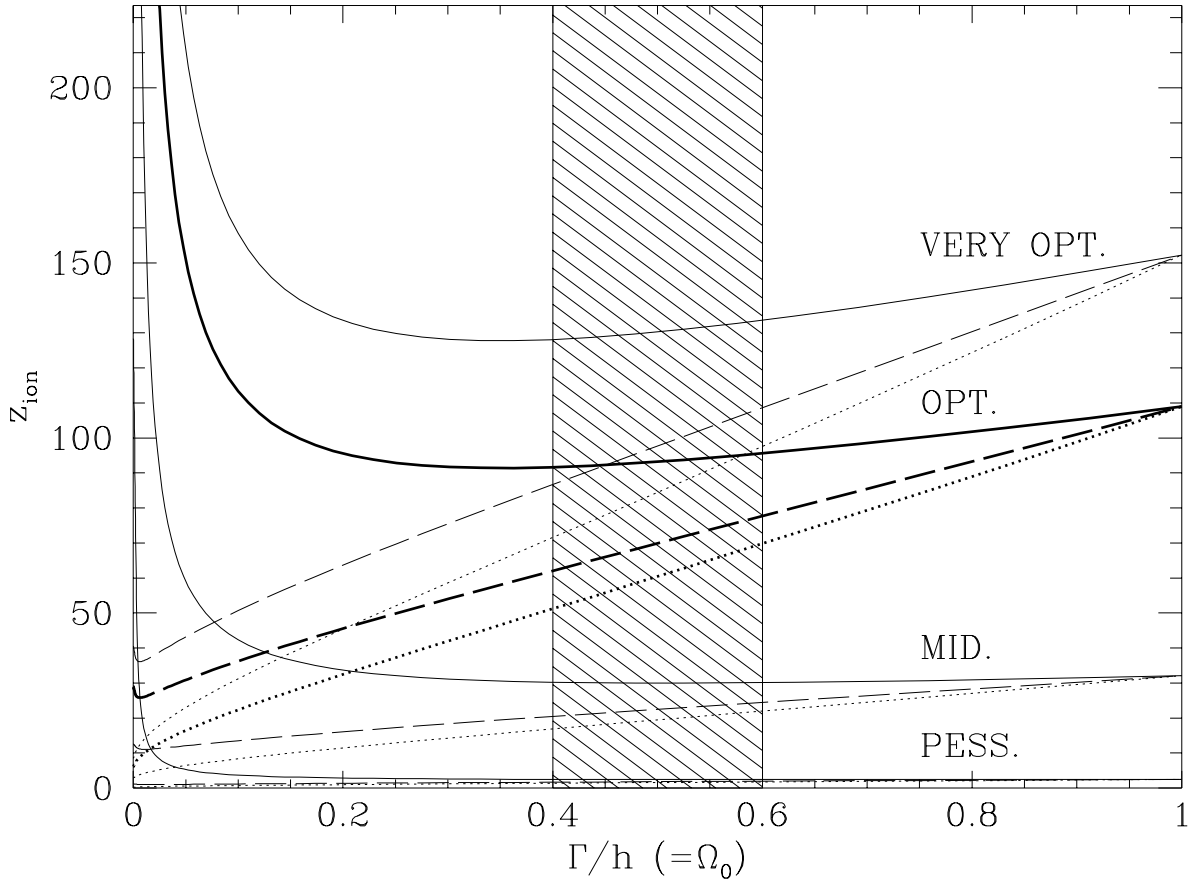}}
\caption{The ionization redshift.}
\label{openreionfig5}
The ionization redshift is plotted for 
the four scenarios described in 
Table~\ref{openreiontable1},
the heavy lines corresponding to the optimistic scenario.
The solid lines are for the open case where $\lambda_0 = 0$.
The dashed lines correspond to the flat case where 
$\lambda_0 = 1-\Omega_0$.
The dotted lines refer to the $\tau$CDM model, where
$\lambda_0 = 0$ and $\Omega_0 = 1$. 
Note that the combination $\Gamma/h$ is really an $h$-independent 
quantity: for the open and flat cases, it is simply equal to 
$\Omega_0$, and for the $\tau$CDM case it depends only on 
$g_*$, the number of 
relativistic particle species.
The vertical shaded region corresponds to values of
$\Gamma$ preferred by power spectrum measurements,
$0.2<\Gamma<0.3$, when $h=0.5$.
\end{figure}

Including the effect of the boost factor, 
equation (9) in
Paper 1
becomes
$$1+\zion = {\sqrt{2}\sigma_8\over\delta_c} 
\>R(h\Oz,r_c,8 h^{-1}\Mpc)
\>B(\Oz,\Lz)\>
\erfc^{-1}\left[{1\over 2\fupp\fion}\right].$$
The ionization redshift $\zion$ is plotted as a function of the shape parameter
in 
Figure~\ref{openreionfig5} 
for the various scenarios specified in 
Table~\ref{openreiontable1}. 
It is seen that for the open model,
the dependence 
on $\Oz=\Gamma/h$ is
typically much weaker than the dependence on other parameters. One reason for
this is that changes in the boost factor and the amplitude 
ratio partially cancel
each other. For $\Lambda$CDM, the $\Oz$-dependence is
stronger, since the boost factor is weaker.
In the  
$\tau$CDM model,
the dependence on $\Gamma/h$
is even stronger, as there is no boost factor whatsoever 
to offset the change in the amplitude ratio. 

\begin{table}
$$
\begin{tabular}{|l|cccc|} 
\hline
                   & Pess.& Mid. & Opt. & Very opt.\\
\hline
$\sigma_8$         & 0.5    & 1    & 1.1 &1.2\\
$\delta_c$         & 2.00   &1.69&1.44&1.33\\
$h$                & 0.5    &0.5&0.8&0.8\\
$M_c [\Ms]$        & $10^8$ &$10^6$&$10^5$&$10^5$\\
$f=\fion\fupp$     & 1      &120&23,000&$10^6$\\
$\erfc^{-1}[1/2f]$ & 0.48   &2.03&3.00&3.55\\
$h^2\Ob$           & 0.010  &0.013&0.015&0.020\\
\hline
\end{tabular}
$$
\caption{Parameters used}
\label{openreiontable1}

\end{table}

The scenarios in Table~\ref{openreiontable1} 
are similar to those in
Paper 1.
In the one labeled ``very optimistic", 
the high value for $\fupp$, 
the net number of produced UV photons per proton, 
is obtained by assuming that the main source of
ionizing radiation is black hole accretion rather than conventional
stars.  Note that this speculative assumption still only increases
$\zion$ by $3.55/3.00 - 1\approx 18\%$, the
efficiency dependence being merely logarithmic.

\begin{figure}[phbt]
\centerline{\epsfxsize=17cm\epsfbox{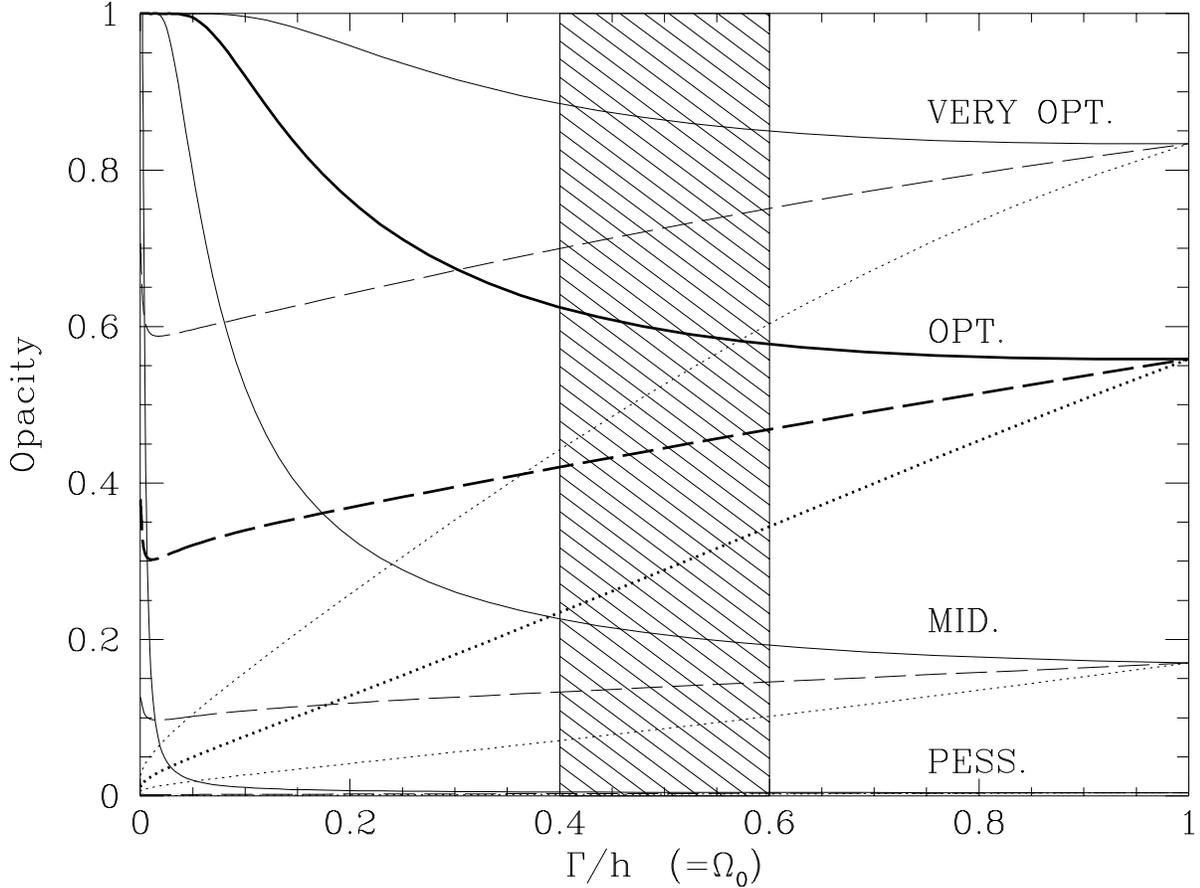}}
\caption{The opacity.}
\label{openreionfig6}
The opacity, 
the probability that a CBR photon is Thomson scattered
at least once since the standard recombination epoch, is
plotted for the four scenarios described in Table 1. 
The solid lines correspond to the open case where $\lambda_0 = 0$.
The dashed lines are for the flat case where
$\lambda_0 = 1-\Omega_0$.
The dotted lines correspond to the $\tau$CDM model, where
$\lambda_0 = 0$ and $\Omega_0 = 1$.
Just as in Figure 5,
note that the combination $\Gamma/h$ is really an $h$-independent
quantity: for the open and flat cases, it is simply equal to
$\Omega_0$, and for the $\tau$CDM case it depends only on
$\gamma$, the number of
relativistic particle species.
The vertical shaded region corresponds to values of
$\Gamma$ preferred by power spectrum measurements,
$0.2<\Gamma<0.3$, when $h=0.5$.
\end{figure}

Figure~\ref{openreionfig6}, 
in a sense the most important plot in this paper, shows the opacity
as a function of $\Gamma/h$ for the 
various scenarios.
Because of the increase in optical depth due to larger lookback times, the 
open model now gives slightly 
larger opacities for lower $\Oz = \Gamma/h$.
However, this dependence is seen to be quite week. 
For $\Lambda$CDM, where the
boost factor contributes less,
the net result is seen to be the opposite; a slight decrease in the opacity for
lower $\Oz = \Gamma/h$. For the $\tau$CDM model, 
where there is neither a boost factor nor an increase in the lookback time, 
this drop in opacity is seen to be much sharper.
Note that the dependence on other uncertain parameters, 
summarized by the four
scenarios in Table~\ref{openreiontable1}, is quite strong. Indeed,
this dependence is stronger than the effect 
of moderate changes in $\Gamma/h$, so in the 
near future, it appears unlikely
that opacity limits will be able to constrain 
the shape parameter except perhaps
in the $\tau$CDM model.

\begin{figure}[phbt]
\centerline{\epsfxsize=17cm\epsfbox{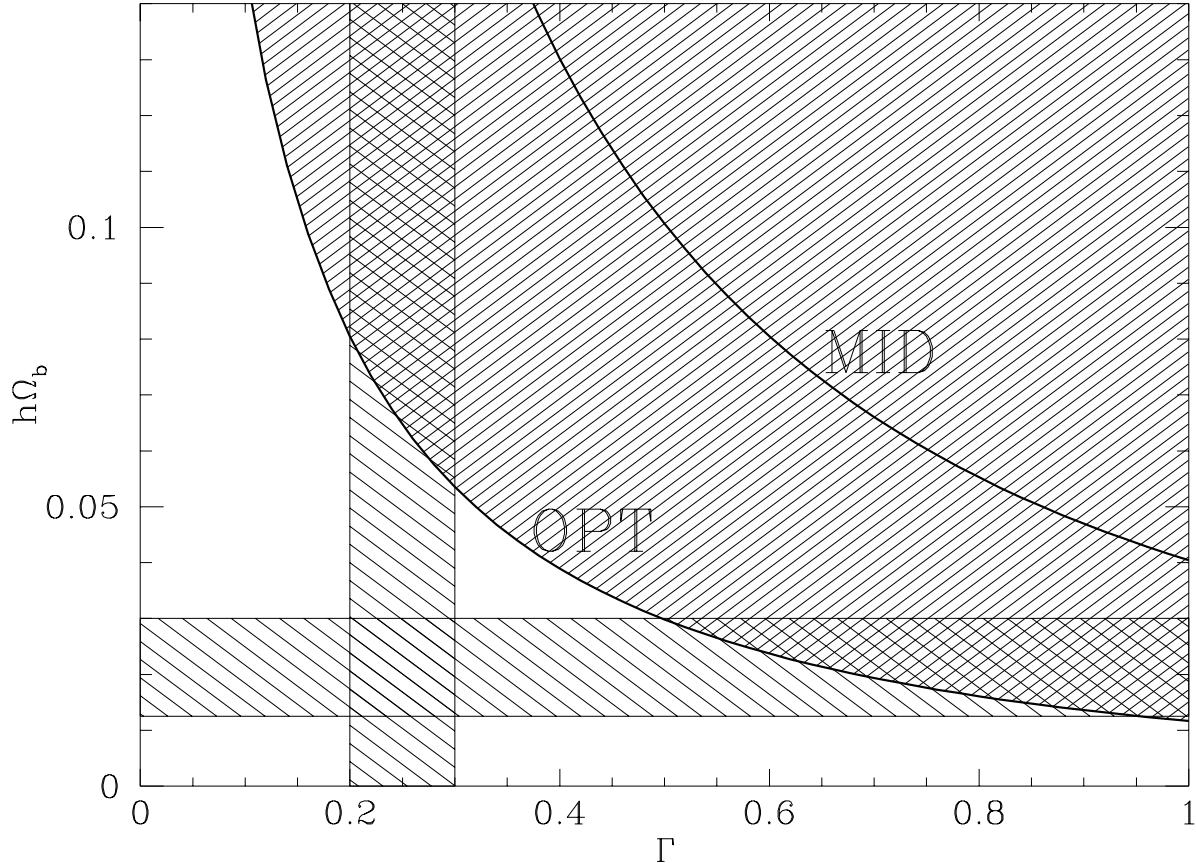}}
\caption{Reionization in $\tau$CDM.}
\label{openreionfig7}
The two 
curves show the baryon density required for $50\%$ opacity in
$\tau$CDM,
{\it i.e.} for reionization
to rescatter $50\%$ of the CBR photons.
The upper and lower heavy curves correspond to the
middle-of-the road and optimistic
scenarios, respectively. Thus even in the optimistic scenario,
$50\%$ opacity cannot be obtained outside of the fine-hatched region.
The horizontal shaded region corresponds to the values of
$h\Omega_b$ allowed by standard nucleosynthesis
($0.01 < h^2\Omega_b < 0.015$) in conjunction with the constraint
$0.5<h<0.8$.
The vertical shaded region corresponds to values of the ``shape
parameter" $\Gamma$ preferred by power spectrum measurements.
Note that these three regions do not intersect.
\end{figure}

The $\tau$CDM situation is summarized 
in 
Figure~\ref{openreionfig7}. 
To attain at least $50\%$ opacity, 
$h\Ob$ must lie above the
heavy curve corresponding to the scenario in question. On the other hand,
nucleosynthesis 
(Smith {\etal} 1993; Walker {\etal} 1991)
places a strict upper bound on this quantity if we assume that
$h\geq 0.5$\footnote{
The $\tau$CDM model also alters the nucleosynthesis process 
(Dodelson, Gyuk \& Turner 1993), but this can only marginally
relax the bounds unless the $\tau$ neutrino is in a 
mass range incompatible with $\tau$CDM (Gyuk \& Turner 1994).}
It is seen that a shape parameter as low as
$\Gamma\simeq 0.25$, which would match large-scale structure observations
(Peacock \& Dodds 1994), is quite
difficult to reconcile with these two constraints. 

\section{Discussion}

Lowering $\Gamma$ is an attractive resolution of the problem 
that arises in reconciling the observed structure in the 
universe on large scales with observations on megaparsec scales. 
The empirical power spectrum is well fit by 
$\Gamma\approx 0.25$ (Peacock \& Dodds 1994). Kamionkowski 
\& Spergel (1993) have found that primordial adiabatic 
fluctuations in an open universe with $\Omega \approx 0.3$ are 
reconcilable with large-scale CMB anisotropy. On degree scales 
Kamionkowski, Spergel \& Sugiyama (1994) require reionization with 
optical depth $\tau\sim 1$ in order to reconcile the low density 
open model with recent experimental limits, if  the lowest 
limits are adopted.
In a low $\Omega_0$ $\Lambda$CDM model, the 
situation is not so critical, but reionization is required 
if the lowest limits (SP91) are adopted on degree 
scales (Gaier {\etal} 1992); 
$\tau\sim 0.5$
suffices however. A similar but slightly more favorable situation 
occurs in a $\tau$CDM model, 
where 
$\Gamma = h\Omega_0 (g_*/3.36)^{-1/2}$ is reduced by increasing 
$g_*$ by a factor of $\sim 4$, 
but some reionization is still required to match SP91.

We have found that reionization giving $\tau$ in the range 0.5 to 1
is readily produced and even natural in open models. 
This is because of the early 
formation of structure in combination with the increased age of the 
universe, effects which compensate for the flattening of 
the power spectrum due to the delay in matter domination.
However, the $\Lambda$CDM and $\tau$CDM models with low $\Omega_0$ 
fare less well in this 
regard, since the loss of small-scale power is not balanced 
by significantly earlier structure formation.
With $\Omega_b$ in the range given by standard nucleosynthesis, 
a significant optical depth $\tau\simgt 0.5$ is 
difficult to attain in either the $\Lambda$CDM or 
$\tau$CDM scenarios.

The authors would like to thank
Wayne Hu, Lloyd Knox, Bernard Sadoulet and Douglas Scott 
for many useful comments. 
This research has been supported in part by a grant from the NSF.


\clearpage

\section{REFERENCES}

\rf Bond, J. R. \& Efstathiou, G. 1984;ApJ;285;L45
 
\rf Carroll, S. M., Press, W. H. \& Turner, E. L. 1992;ARA\&A;30;499

\rf Crittenden, R., Davis R. \& 
Steinhardt P. J. 1993;ApJ;417;L13

\rf Dodelson, S., Gyuk, G. \& Turner, M. S. 1994a;
Phys. Rev. Lett.;72;3754

\rf Dodelson, S., Gyuk, G. \& Turner, M. S. 1994b;
Phys. Rev. D;49;5068

\rf Efstathiou, G., Bond, J. R. \& White, S. D. M. 1992;MNRAS;258;1P

\rf Feynman, R. P. 1939;Phys. Rev.;56;340

\rf Gaier {\etal} 1992;ApJ;398;L1
 
\rn Gyuk, G. \& Turner, M. S. 1994,
astro-ph/9403054 preprint.

 
\rf Gorski, K. M., Stompor, R \& Juszkiewicz, R. 1993; ApJ Lett;410;L1

\rf Gunn, J. E. \& Peterson, B. A. 1965;ApJ;142;1633

\rf Kamionkowski, M. \& Spergel, D. N. 1994;ApJ;432;7

\rf Kamionkowski, M., 
Spergel, D. N. \& Sugiyama, N. 1994;ApJ;426;L57
 
\rn Kolb, E. \& Turner, M. S. 1990, ``The Early Universe",
Addison-Wesley
 
\rf Peacock, J. A. \& Dodds, S. J. 1994;MNRAS;267;1020
 
\rn Peebles, P. J. E. 1980, 
{\it The large-scale structure of the universe},
Princeton U. P., Princeton.

\rf Smith, M. S., Kawano, L. H. \& Malaney, R. A.
1993;ApJS;85;219

\rf Steidel, C. C. \& Sargent, W. L. W. 1987;ApJ (Letters);318;L11
 
\rf Sugiyama, N., Silk, J. \& Vittorio, N. 1993; ApJ Lett;419;L1

\rf Tegmark, M., Silk, J. \& Blanchard, A. 1994; ApJ;420;484
 
 
\rf Vittorio, N. \& Silk, J 1992; ApJ Lett;385;L9
 
\rf Walker, P. N. {\etal} 1991;ApJ;376;51

\rf Webb, J. K., Barcons, X., Carswell, R. F., \&
Parnell, H. C. 1992;MNRAS;255;319

\rf Wollack, E.J. {\etal} 1993; ApJ;419;L49


\clearpage

\clearpage

\clearpage

\clearpage

\clearpage

\clearpage

\clearpage

\end{document}